\documentclass{aa}  

\usepackage{graphicx}
\usepackage{txfonts}
\usepackage{lipsum}
\usepackage{subcaption}         % necessary for continued figures, example in section 3
\usepackage{lscape}             % to rotate a single page table, example in appendix.
\usepackage{placeins}           % useful with \FloatBarrier, to keep 
\newcommand{\fcyl}{f_\text{cyl}}
\usepackage[breaklinks=true]{hyperref}
\hypersetup{colorlinks=true,allcolors=blue}
\begin{document}

   \title{Bonnor-Ebert sphere collapse in filamentary structures}

   \subtitle{}

   \author{S. Heigl\inst{1,2}
          \and
          A. Burkert\inst{1,2,3}
          }

   \institute{Universit\"ats-Sternwarte, Ludwig-Maximilians-Universit\"at M\"unchen,
             Scheinerstr. 1, 81679 Munich, Germany\\
             \email{heigl@usm.lmu.de}
         \and
             Excellence Cluster ORIGINS,
             Boltzmannstrasse 2, 85748 Garching, Germany
         \and
             Max-Planck Institute for Extraterrestrial Physics,
             Giessenbachstr. 1, 85748 Garching, Germany
             }

   \date{Received; accepted}

% \abstract{}{}{}{}{}
% 5 {} token are mandatory

  \abstract
   {Star formation within filaments may arise due to the growth of cores according
   to linear perturbation theory. This implies a minimum core separation, as
   shorter modes would not be able to grow. While many observations agree with core
   separations by theoretical predictions, some observations also show star forming
   cores which lie closer together than the minimum wavelength given by perturbation
   theory.}
   {We explore whether non-linear effects during the late stages of core growth can explain
   the discrepancy between theory and observations.}
   {We perform three-dimensional hydrodynamical simulations with the \textsc{ramses}
   code to follow the evolution of initial perturbations within filaments and compare
   the measured growth rates to expectations from theoretical models.}
   {Non-linear evolution sets in as soon as the core mass reaches a value where
   the gravitational potential is not any longer dominated by the cylindrical potential
   of the filament but by the spherical potential of the Bonnor-Ebert sphere. Consequently,
   core collapse is not triggered by the loss of hydrostatic stability of the filament but by
   the loss of hydrostatic stability of the Bonnor-Ebert sphere. As the core is embedded in
   the filament, the maximum core mass is given by the pressure within the filament which
   results in a constant line-mass threshold for core collapse.}
   {As core collapse is triggered as soon as overdensities reach a certain line-mass,
   cores which form as large line-mass perturbations during filament formation can go
   into direct collapse even if their separation is closer than predicted by linear
   perturbation theory. Therefore, our result can explain the discrepancy between theory
   and observations.}


   \maketitle
   \nolinenumbers

%%%%%%%%%%%%%%%%%%%%%%%%%%%%%%%%%%%%%%%%%%%%%%%%%%%%%%%%%%%%%%
\section{Introduction}

  Star formation is a multi-scale process. Relatively defuse gas within molecular
  clouds of sizes of a few tens of parsecs is gravitationally compressed to high
  densities within star-forming cores with sizes of a few tenths of parsecs.
  However, this gravitational cascade has an intermediate step forming filamentary
  structure containing regularly spaced cores already seen in early extinction
  and infrared observations \citep{schneider1979}. The importance of filaments as
  initial condition for core formation was emphasised by high dynamic range dust
  observations of the \textit{Herschel Space Telescope} \citep{andre2010,
  molinari2010}. It revealed a high degree of filamentary organisation in nearby
  molecular clouds and established a clear connection between cores and filaments
  \citep{konyves2015}. Newer continuum and line observations have refined this
  picture and filamentary structure is now well established at all scales in the
  interstellar medium of the Milky Way (see \citet{hacar2023} for a recent review).

  A fundamental question for the formation of stars has been the influence of
  filaments on the core collapse process. In principle, filaments are able to affect
  the distribution and masses of stars by providing the gravitational environment
  and mass reservoir for star-forming cores. The idea of filament fragmentation into
  quasi-regular spaced cores that eventually collapse and form stars has been
  predicted by theoretical models (see Sect.~\ref{sec:theory}) with a minimum
  wavelength required for perturbations to grow, called the critical wavelength,
  and a fastest growing mode of about two times the minimum wavelength, also called
  the dominant wavelength. Although cores in principle can form on any wavelength
  during filament formation, non-growing modes are not expected to become massive
  enough to eventually collapse. While the comparison of observations and theory
  is not straightforward and there are issues concerning low number statistics,
  many studies do find separations agreeing with the fastest growing mode, with
  or without considering a non-isothermal contribution \citep{jackson2010,
  miettinen2012, busquet2013, lu2014, wang2014, beuther2015, contreras2016,
  kainulainen2016, liu2018, lu2018, cao2022, he2023, chung2024, yang2024}.
  Nevertheless, there are also studies which show closer separations, either two
  times shorter, as would be the case for the critical wavelength, or even closer
  by a factor of three or more, as would be the case for the three dimensional
  Jeans length \citep{hartmann2002, tafalla2015, henshaw2016, williams2018,
  zhou2019, konyves2020, zhang2020, hwang2026}.

  Theoretically, there are many processes which can influence core spacing.
  Stabilising effects such as a stiffer equation of state \citep{gehman1996a,
  mclaughlin1996, hosseinirad2018, coughlin2020} and magnetic fields
  \citep{nagasawa1987, gehman1996b, fiege2000, hosseinirad2017,
  hanawa2019} generally increase core spacing. Accretion \citep{clarke2016},
  geometrical displacement \citep{gritschneder2017} and central hubs
  \citep{zhen2025} can completely dominate over linear perturbation theory.
  Moreover, the longitudinal collapse of filaments can either form dominant
  cores at the end of filaments or move existing cores closer together over time
  \citep{bastien1983, burkert2004, pon2012, clarke2015, heigl2022, miettinen2022,
  hoemann2023a}. There are also studies which argue for a more complex, hierarchical
  fragmentation where the separation at low densities follows the predicted mode
  of filament fragmentation while having separations similar to the local Jeans
  length at high densities \citep{kainulainen2013, takahashi2013, teixeira2016,
  kainulainen2017, mattern2018,palau2018, svaboda2019}.

  A related problem is the question of what fundamentally drives
  the collapse of cores within filaments. Isolated cores are expected to
  collapse at a maximum mass, or density contrast respectively, given by the
  Bonnor-Ebert solution \citep{ebert1955, bonnor1956}. The predicted profile
  for an isothermally stable Bonnor-Ebert sphere is also observed in isolated
  cores \citep{alves2001, tafalla2004}. However, cores are usually not truncated
  which raises the question whether the Bonnor-Ebert sphere is a realistic
  representation of real cores \citep{shu1977, vazquez2005, vazquez2019}. Recent
  numerical studies argue that cores are limited by their effective tidal radius
  and that their internal turbulent pressure has to be taken into account for
  determining its stability \citep{moon2024, moon2025}. The question of stability
  of cores within filaments is even more complex. Similar to the Bonnor-Ebert
  sphere, filaments also have an upper limit of how much mass they can support
  in hydrostatic equilibrium. Therefore, core collapse may also be triggered
  once the line-mass exceeds this upper mass limit.

  In this paper, we explore what drives the collapse of cores in idealised
  filaments and whether this has implications on the expected separation. In
  Sect.~\ref{sec:concepts} we discuss the basic physical principles of hydrostatic
  equilibrium solutions and perturbation theory in filaments. We describe our
  numerical set-up in Sect.~\ref{sec:setup}. Our results on core collapse and
  core separations are given in Sect.~\ref{sec:results} which we summarise and
  discuss the implication in Sect.~\ref{sec:conclusion}.

%%%%%%%%%%%%%%%%%%%%%%%%%%%%%%%%%%%%%%%%%%%%%%%%%%%%%%%%%%%%%%
\section{Basic concepts}
  \label{sec:concepts}

  \subsection{Hydrostatic isothermal solutions}

  In general, self-gravitating, hydrostatic solutions are found by inserting the
  hydrostatic condition
  \begin{equation}
    \frac{\nabla P}{\rho} = - \nabla\Phi,
  \end{equation}
  with the density $\rho$, pressure $P$ and the gravitational potential $\Phi$,
  into the Poisson equation for gravity
  \begin{equation}
    \Delta \Phi = 4\pi G\rho,
  \end{equation}
  with $G$ being the gravitational constant. In the specific case of isothermality, the
  equation can be rewritten to
  \begin{equation}
    \Delta_\xi \psi = \exp(-\psi),
    \label{eq:iso}
  \end{equation}
  using a variable transform for the potential $\Phi = \psi c_s^2$ and the radius
  $r^2 = \frac{c_s^2}{4\pi G\rho_c}\xi^2$. Here, $c_s$ is the isothermal sound speed
  and $\rho_c$ is the profile's central density. $\Delta_\xi$ is the Laplace operator
  with respect to $\xi$ and depends on the dimensionality of the profile.

  For the cylindrical profile in two dimensions, Eq.~\ref{eq:iso} transforms to
  \begin{equation}
    \frac{1}{\xi}\frac{d}{d\xi}\left(\xi\frac{d\psi}{d\xi}\right) = \exp(-\psi),
    \label{eq:fil}
  \end{equation}
  which has an analytical solution given by the isothermal hydrostatic cylinder
  \citep{stodolkiewicz1963, ostriker1964}:
  \begin{equation}
    \rho(r) = \frac{\rho_c}{\left(1+\left(r_\text{cyl}/H\right)^{2}\right)^{2}}
    \label{eq:rho}
  \end{equation}
  where $r_\text{cyl}$ in this case is the cylindrical radius and the radial scale height $H$
  is given by the term:
  \begin{equation}
     H^2 = \frac{2c_s^2}{\pi G \rho_c}.
     \label{eq:h}
  \end{equation}
  This profile also can only support a maximum mass in hydrostatic equilibrium,
  in this case per unit length. It is calculated by integrating the profile radially
  to infinity:
  \begin{equation}
     \left(\frac{M}{L}\right)_\text{crit} = \frac{2c_\text{s}^2}{G}
     \label{eq:lmcrit}
  \end{equation}
  In the case of pressure truncation at the boundary density $\rho_b$ at radius $R$,
  one can define the parameter $\fcyl$ which gives the value of the current line-mass
  compared to the critical value \citep{fischera2012}:
  \begin{equation}
    \fcyl = \left(\frac{M}{L}\right)/\left(\frac{M}{L}\right)_\text{crit}
    = \frac{1}{1 + \left(H/R\right)^2}.
  \end{equation}
  The radius is then given by:
  \begin{equation}
    R = H\left(\frac{\fcyl}{1-\fcyl}\right)^{1/2}
  \end{equation}
  and the central density is set by the boundary density as:
  \begin{equation}
    \rho_c = \frac{\rho_b}{(1-\fcyl)^2}.
    \label{eq:contrast}
  \end{equation}

  For the spherical profile in three dimensions, Eq.~\ref{eq:iso} transforms to the
  isothermal Lane-Emden equation
  \begin{equation}
    \frac{1}{\xi^2}\frac{d}{d\xi}\left(\xi^2\frac{d\psi}{d\xi}\right) = \exp(-\psi),
    \label{eq:le}
  \end{equation}
  which has a well-known solution: the Bonner-Ebert sphere \citep{ebert1955,
  bonnor1956}. While the profile follows $r^{-2}$ at large radii, it can only
  be calculated numerically. The Bonnor-Ebert sphere has a fixed upper mass
  limit it can support which, in contrast to a filament, depends on the
  isothermal external pressure $P_\text{ext} = \rho_\text{ext}c_s^2$ and is
  given by:
  \begin{equation}
    M_\text{BE} = \frac{1.18 c_s^4}{P_\text{ext}^{1/2} G^{3/2}} = \frac{1.18 c_s^3}{\rho_\text{ext}^{1/2} G^{3/2}}
    \label{eq:Mbe}
  \end{equation}
  which it reaches at a density contrast of $\rho_c/\rho_b = 14.1$.

  \subsection{Filament perturbation theory}
  \label{sec:theory}

  \begin{figure*}
    \sidecaption
    \includegraphics[width=12cm]{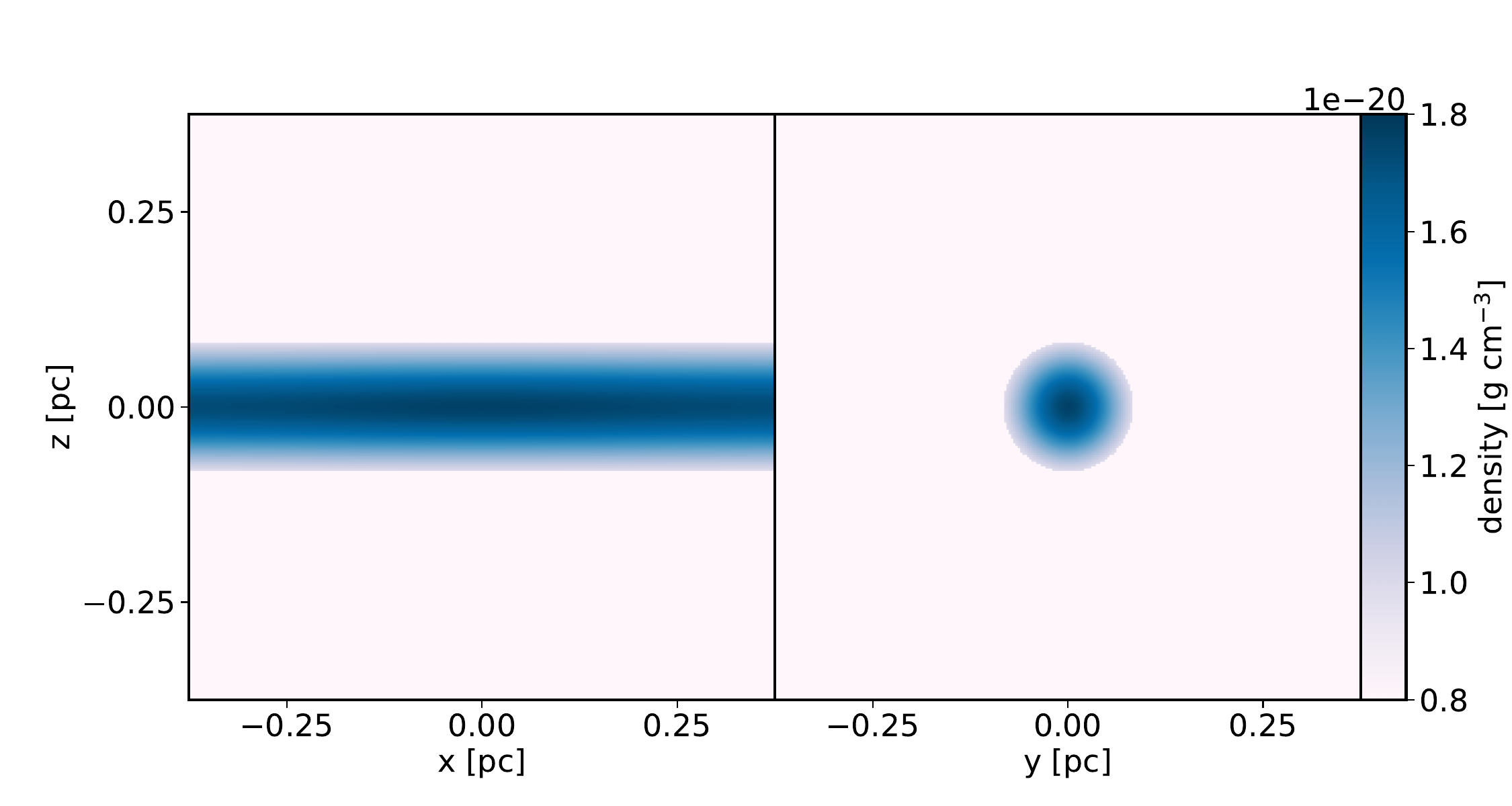}
    \caption{Example of the filament orientation within the simulation box. Shown here on the
    left hand side is a slice along the filament axis at $y=0.0\,\mathrm{pc}$ and a slice
    perpendicular of the filament at $x=0.0\,\mathrm{pc}$ on the right hand side. This
    particular filament has a line-mass of $\fcyl=0.25$ and a perturbation on the dominant
    wavelength which is also the size of the box.}
    \label{fig:slice}
  \end{figure*}

  Small perturbations to the variable $y$ of size $\epsilon$ along the axis $x$ of
  hydrostatic isothermal cylinders with a form
  \begin{equation}
    y(x, t) = y_0 + y_1(x, t) = y_0\left(1 + \epsilon\exp(ikx - i\omega t)\right)
    \label{eq:evo}
  \end{equation}
  will grow for wavelengths $\lambda = 2\pi/k$ which are greater than a critical
  wavelength on a timescale $\tau = 1/\omega$ \citep{stodolkiewicz1963}. Perturbations
  with separations shorter than the critical wavelength will lead to oscillating
  soundwaves within the filament. While many observed cores on shorter wavelengths
  could be the result of the formation of the filament, these cores should in
  principle not be able to grow in mass and form cores. The growth timescale has
  a minimum value for separations which are about double the critical wavelength,
  also called the dominant wavelength \citep{nagasawa1987}. While perturbations
  can grow on any wavelength longer than the critical one, the fastest growing
  mode is expected to dominate for small initial perturbations.

  Both, the critical as well as the dominant wavelength, depend on the line-mass of
  the filament. For a value of $\fcyl=1.0$, the critical wavelength is about
  $\lambda_\text{crit}=3.94 H$ and the dominant wavelength is about
  $\lambda_\text{dom}=7.82 H$ with lower values for lower
  line-masses \citep{stodolkiewicz1963, larson1985, nagasawa1987, inutsuka1992}.
  \citet{nagasawa1987} determined the length- and timescales for several other values of
  $\fcyl$ which where interpolated by \citet{fischera2012}. We use this approximation in
  order to determine the fastest growing mode and its growth timescale.

  \citet{fischera2012} also determined for which values of the filament line-mass, the forming
  core exceeds the Bonnor-Ebert mass if all the mass on the dominant wavelength is accumulated
  into the core. This is done by equating the total mass of the dominant wavelength
  \begin{equation}
    M_\text{dom} = \lambda_\text{dom}(\fcyl) \fcyl \left(\frac{M}{L}\right)_\text{crit}
  \end{equation}
  to Eq.~\ref{eq:Mbe}. This shows that if cores are forming on the dominant wavelength, only
  cores forming in filaments of line-masses $0.20 < \fcyl < 0.89$ will be able to collapse, a
  result we will be testing with our simulations.

%%%%%%%%%%%%%%%%%%%%%%%%%%%%%%%%%%%%%%%%%%%%%%%%%%%%%%%%%%%%%%
\section{Numerical set-up}
  \label{sec:setup}

  Our simulations were run with the code \textsc{ramses} \citep{teyssier2002} which employs
  a second-order Godunov scheme on a Cartesian grid in order to solve the Euler equations
  in their conservative form. As solver we chose the MUSCL scheme
  \citep[Monotonic Upstream-Centred Scheme for Conservation Laws,][]{vanLeer1977} together
  with the HLLC-Solver \citep[Harten-Lax-van Leer-Contact,][]{toro1994} and the multidimensional
  MC slope limiter \citep[monotonized central-difference,][]{vanLeer1979}.

  As we study the growth of perturbations within filaments, we set up density profiles according
  to Eq.~\ref{eq:rho} with varying line-masses. The filament axis is always aligned with the x-axis
  and lies in the centre of the y-z plane. In order to minimise accretion onto the filament, the
  background density is always given by a very low particle density of $2.5 \,\mathrm{cm^{-3}}$,
  corresponding to a volume density of $9.87\times 10^{-24}\,\mathrm{g\,cm^{-3}}$ for a molecular
  weight of $\mu = 2.36$, and the contrast to the filament boundary density is set to a value of
  $10^3$. Therefore, the boundary density is always given by $2.5 \times 10^3\,\mathrm{cm^{-3}}$,
  or $9.87\times 10^{-21}\,\mathrm{g\,cm^{-3}}$, and the central density is set according to
  Eq.~\ref{eq:contrast}. For values above the boundary density, we use an isothermal equation of
  state with a temperature of $10.0\,\mathrm{K}$. If the density falls below the boundary density,
  we apply an isobaric equation of state in order to mimic a constant pressure environment with
  a value of $p_{ext} = \rho_b c_s^2$, where $c_s$ is the isothermal sound speed. An example of
  the filament orientation is given in Fig.~\ref{fig:slice} where we show two density slices of
  the initial condition for a line-mass of $\fcyl=0.25$. The left hand side shows a slice along
  the filament axis at $y=0.0\,\mathrm{pc}$ and the right hand side shows a slice perpendicular
  to the filament axis at $x=0.0\,\mathrm{pc}$.

  \begin{figure}
    \centering
    \includegraphics[width=1.0\columnwidth]{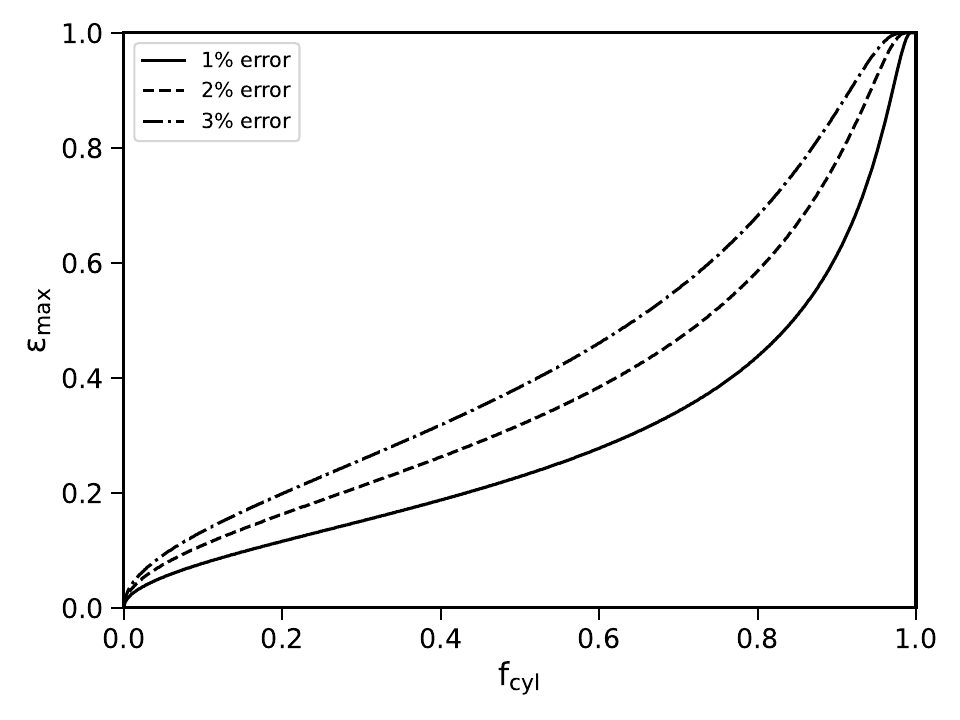}
    \caption{Maximum perturbation strength in dependence of the line-mass in order to keep the
    relative line-mass error below different thresholds. The thresholds plotted are 1, 2 and
    3\% given by the solid, dashed and dashed-dotted lines, respectively.}
    \label{fig:lmerror}
  \end{figure}

  On top of the density profile we superimpose a perturbation as defined in Eq.~\ref{eq:evo}
  of the form
  \begin{equation}
    \rho_c(x, t=0.0) = \rho_{c0}\left(1+\epsilon\cos(2\pi x/ \lambda)\right)
    \label{eq:rhoini}
  \end{equation}
  and adapt the scale height in Eq.~\ref{eq:h} accordingly. In addition, we include an initial
  velocity which we determine by using Eq.~\ref{eq:rhoini} in the continuity equation
  \begin{equation}
    \frac{\partial\rho}{\partial t} - \nabla(\rho v_x) = 0
  \end{equation}
  which results in a velocity perturbation of
  \begin{equation}
    v_{x1}(x, t=0.0) = -\frac{\epsilon\lambda}{2\pi\tau}\sin(2\pi x/ \lambda)
  \end{equation}
  As one can see, the velocity is shifted by a quarter wavelength compared to the density
  perturbation and points towards the overdensity. Because of the condition of a constant
  pressure environment and the relation between the central and boundary density set by
  Eq.~\ref{eq:contrast}, the density perturbation leads to a non-sinusoidal perturbation
  in the line-mass where the line-mass maximum is given by a line-mass perturbation
  strength of
  \begin{equation}
    \epsilon^{\text{max}}_f = \frac{1-f_0}{f_0}\left(1-\sqrt{\frac{1}{1+\epsilon}}\right)
    \label{eq:eps}
  \end{equation}
  and the line-mass minimum by a line-mass perturbation strength of
  \begin{equation}
    \epsilon^{\text{min}}_f = \frac{1-f_0}{f_0}\left(1-\sqrt{\frac{1}{1-\epsilon}}\right).
  \end{equation}
  As a result, the line-mass is not strictly conserved. However, even for large values of
  $\epsilon$ this only introduces a small relative error in line-mass. We visualise this in
  Fig.~\ref{fig:lmerror} where the maximum perturbation strength $\epsilon_\text{max}$ is shown
  in order to stay below a given relative error threshold in line-mass. These are given by the
  solid, dashes and dashed-dotted lines which show the threshold of 1, 2 and 3\% relative error,
  respectively. The values of $\epsilon_\text{max}$ are much larger than the initial perturbation
  strengths we typically use in our simulations, which are around 0.01 and less. Therefore, this
  effect only plays a minor role as even the error introduced by the discretisation of the filament
  on the grid can easily reach 1\%.

  The boundary conditions of the simulation box are set to periodic along the filament axis and
  to open boundaries in the y- and z-direction. The periodic condition prevents the filament of
  collapsing along its axis and allows us to study the growth of density perturbations without
  the longitudinal collapse or the formation of end cores. This is also the case for
  filaments in the interstellar medium, if they have shallow density gradients at their ends
  \citep{hoemann2023b}. The open boundary conditions allow inflow into the box which over time can
  lead to artifacts in the ambient medium. Therefore, we define a cylindrical shell with a radius
  of half the box size in which the density is constantly reset to the background density and the
  velocity is reset to zero.

  We use adaptive mesh refining with the minimum and maximum resolution of $128^3$ and
  $512^3$ cells, respectively. As refinement threshold we use the Truelove criterion where
  the Jeans length has to be resolved by 128 cells \citep{truelove1997}. The relatively
  high number of cells guarantees that the central region of the filament is resolved on
  the highest level which is more than sufficient to resolve the growing core to a very
  high degree while leaving the ambient medium at low resolution. We do not use any sink
  particle formation but stop the simulation as soon as the Jean's length at our highest
  resolution is not resolved by at least four cells which indicates the runaway collapse.

  %%%%%%%%%%%%%%%%%%%%%%%%%%%%%%%%%%%%%%%%%%%%%%%%%%%%%%%%%%%%%%
\section{Simulation results}
  \label{sec:results}

  Here, we discuss the results of our simulations. In the first subsection, we will clarify
  at which point the Bonnor-Ebert sphere influences the core growth and collapse. In the
  second subsection, we explore whether the Bonnor-Ebert sphere has an effect on the
  separation of collapsing cores.

  \subsection{Non-linear evolution}
  We systematically explore the core growth on the dominant wavelength for different values
  of the filament line-mass in order to measure any deviation. Therefore, we set the box
  size to the dominant wavelength for a given line-mass calculated from the interpolation
  of \citet{fischera2012} and add a single perturbation with the length of the box size.
  The initial perturbation strength is set to $\epsilon = 0.01$ for low line-masses but is
  reduced for higher line-masses in order to give the simulation more time to evolve. For
  very high line-masses, the dominant wavelength becomes so small that the filament radius
  would exceed the simulation box. In this case, we double the box size and add a perturbation
  with a length of half the box size, effectively following the evolution of two peaks on
  the dominant wavelength.

  \begin{figure}
    \centering
    \includegraphics[width=1.0\columnwidth]{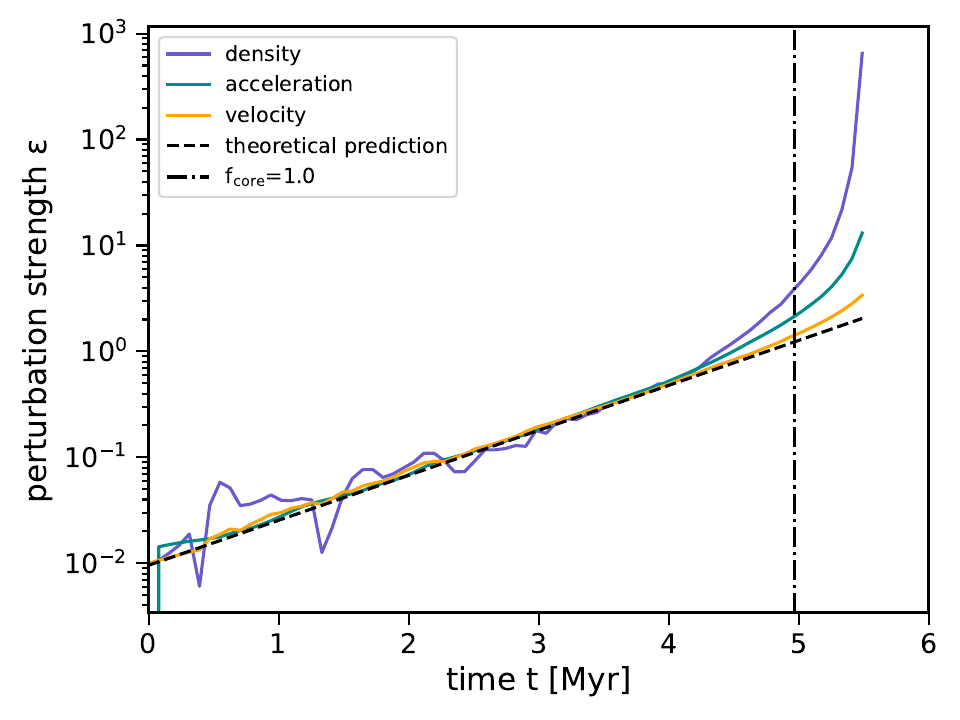}
    \caption{Time evolution of the perturbation strength of different variables, shown
    as solid lines, compared to the theoretical prediction, given by the dashed line.
    The blue, cyan and orange lines show the evolution of the maximum in density,
    x-acceleration and x-velocity, respectively, for a filament of a line-mass of
    $\fcyl=0.25$.}
    \label{fig:perttime}
  \end{figure}

\begin{figure}
    \centering
    \includegraphics[width=1.0\columnwidth]{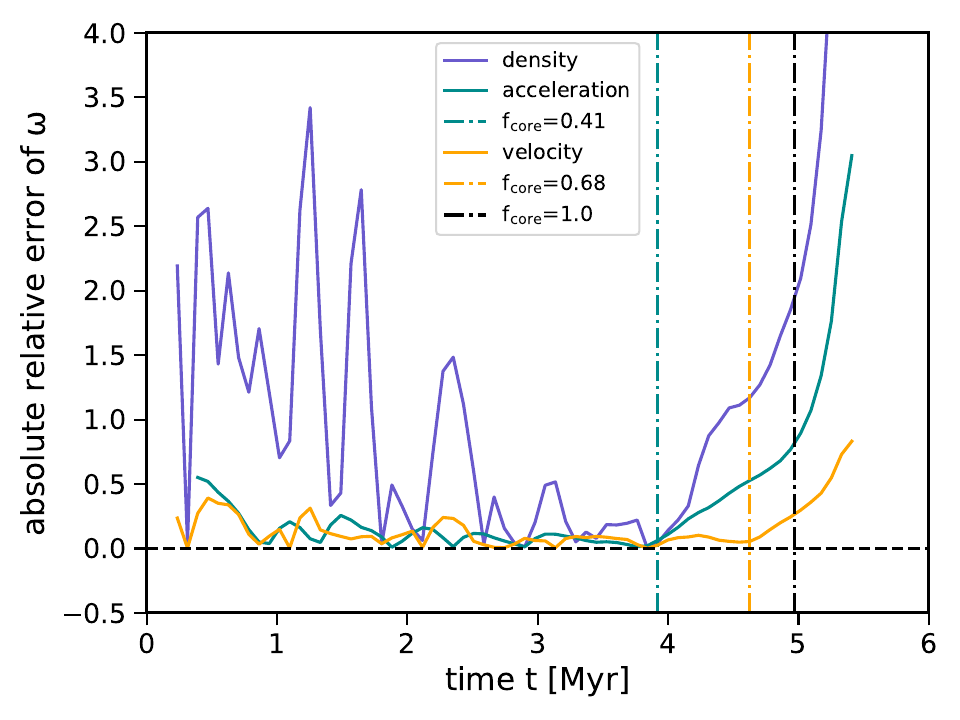}
    \caption{Time evolution of the absolute relative error of the growth time scale
    $\omega$ measured in the simulation for a line-mass of $\fcyl=0.25$. The solid
    blue, cyan and orange line shows the error of the density, x-acceleration and
    x-velocity, respectively. We define the begin of the non-linear evolution by
    determining the point in time when the error exceeds a value of 5\% when measuring
    back from the end of the simulation. The thresholds are indicated by the vertical
    dashed-dotted lines, for which we also determine the line-masses at the position
    of the core which are given in the legend.}
    \label{fig:errtime}
  \end{figure}

  In Fig.~\ref{fig:perttime} we show the time evolution of the perturbation strength of
  different variables in a filament with a line-mass of $\fcyl=0.25$. The blue, cyan and
  orange lines show the strength of the measured maximum in density, x-acceleration and
  x-velocity, respectively. All perturbed values follow the theoretical prediction shown
  as dashed black line and exhibit a strong exponential growth at some point, however the
  onset of the deviation is not the same. In order to visualise this more clearly, we
  determine the absolute relative error of the measured growth rate to the theoretical value
  of $\omega$. We determine the growth rate in the simulation by taking the time gradient of
  the logarithm of Eq.~\ref{eq:evo} where the unperturbed value is substracted first. For
  variables which unperturbed value is zero, such as the acceleration or the velocity, this
  method has the advantage of not depending on the initial magnitude of the perturbation.
  As taking the time gradient on an output to output basis introduces strong noise, we first
  smooth our data by taking a rolling average over five outputs of our simulation. The
  result is given in Fig.~\ref{fig:errtime} where the colours show the same variables as
  above. All variables show wave-like patterns in their evolution due to the discetisation
  of the initial condition leading to an oscillation in the relative error which dampens
  over time. One can see that the onset of the non-linear evolution is the same for the
  density and acceleration but is delayed for the velocity.

  In order to determine it the respective thresholds in time and line-mass, we measure
  back from the exponential growth at end of simulation and determine when the absolute
  relative error exceeds a value of 5\%. As the density and the acceleration show the
  same threshold, we use the acceleration as it systematically has a smaller error. Both
  threshold values, one for the acceleration and one for the velocity, are shown in
  Fig.~\ref{fig:errtime} as dashed-dotted vertical lines with the respective colour and
  the corresponding line-mass value at the position of the core is given in the legend.
  The fact that the non-linear evolution begins before the critical line-mass is reached,
  shows that it is not the loss of radial hydrostatic equilibrium of the filament
  determining the core evolution or even the collapse. Moreover, the fact that there
  are two separate thresholds for the acceleration and velocity shows that there seems
  to be three separate phases in the core evolution. In the first phase, the growth
  follows linear perturbation theory. In the second phase, the density, and thus
  the acceleration, grows faster than expected without the velocity changing its behavior.
  This means there is a process which concentrates the density in the core. In the third
  phase, the velocity shows an exponential deviation. This means that the core collapse
  has set in and the velocity is growing according to the free-fall collapse.

\begin{figure}
    \centering
    \includegraphics[width=1.0\columnwidth]{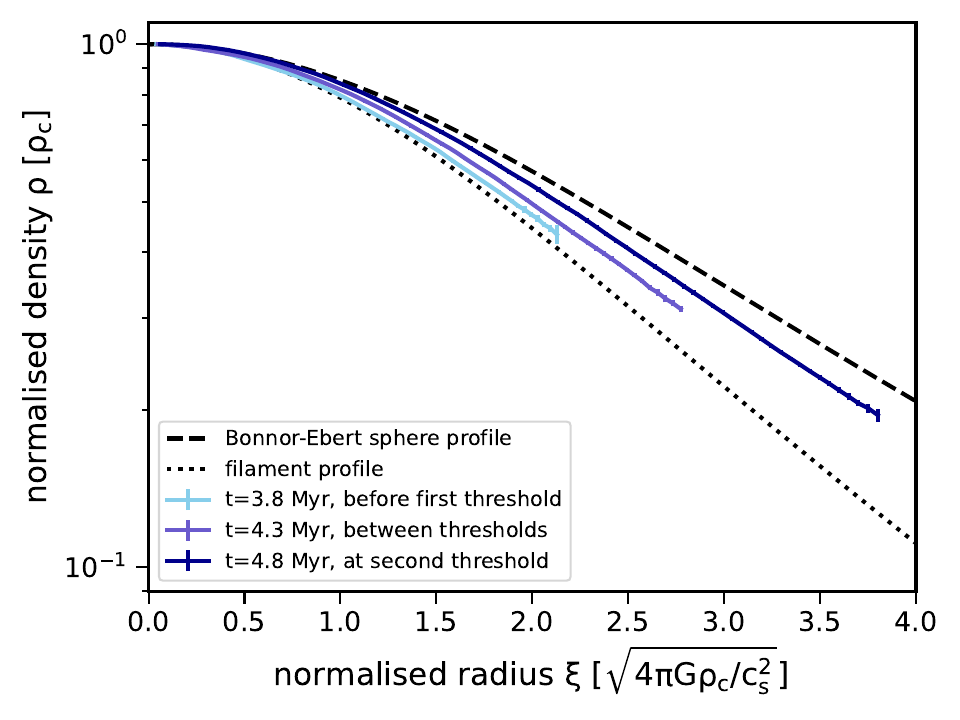}
    \caption{Evolution of the radial density profile of the core from the linear
    growth phase until the start of the collapse for a line-mass of $\fcyl=0.25$.
    Shown are snapshots before the first threshold in light blue, between the first
    and second threshold in blue and just after the second threshold in dark blue.
    The density as well as the radius are normalised to a dimensionless form in
    order to compare profiles with different concentrations. The filament profile
    as defined in Eq.~\ref{eq:rho} is given as dotted line and the Bonnor-Ebert sphere
    profile as dashed line.}
    \label{fig:profile}
  \end{figure}

  This effect can also be seen in the evolution of the radial density profile over
  time. In Fig.~\ref{fig:profile} we show snapshots of the profile in the three
  different phases for the filament with a line-mass of $\fcyl=0.25$. The density
  as well as the radius are normalised to a dimensionless form. This means we can
  compare the density profiles directly to the general filament profile as defined
  in Eq.~\ref{eq:rho}, given by the dotted line, and the Bonnor-Ebert sphere profile,
  shown as dashed line. In the first phase plotted as the light blue line, the
  density profile follows the filament profile. As soon as the filament reaches
  the second phase as given by the blue line, the density gets more concentrated
  and transitions to the Bonnor-Ebert sphere profile. The last snapshot in dark
  blue shows the density profile just after the second threshold. Here, the profile
  resembles a Bonnor-Ebert sphere. For later times, the density profile maintains
  the shape of a Bonnor-Ebert sphere as is expected during collapse
  \citep{larson1969, penston1969, whithworth1985, foster1993, keto2010, naranjo2015}.

\begin{figure}
    \centering
    \includegraphics[width=1.0\columnwidth]{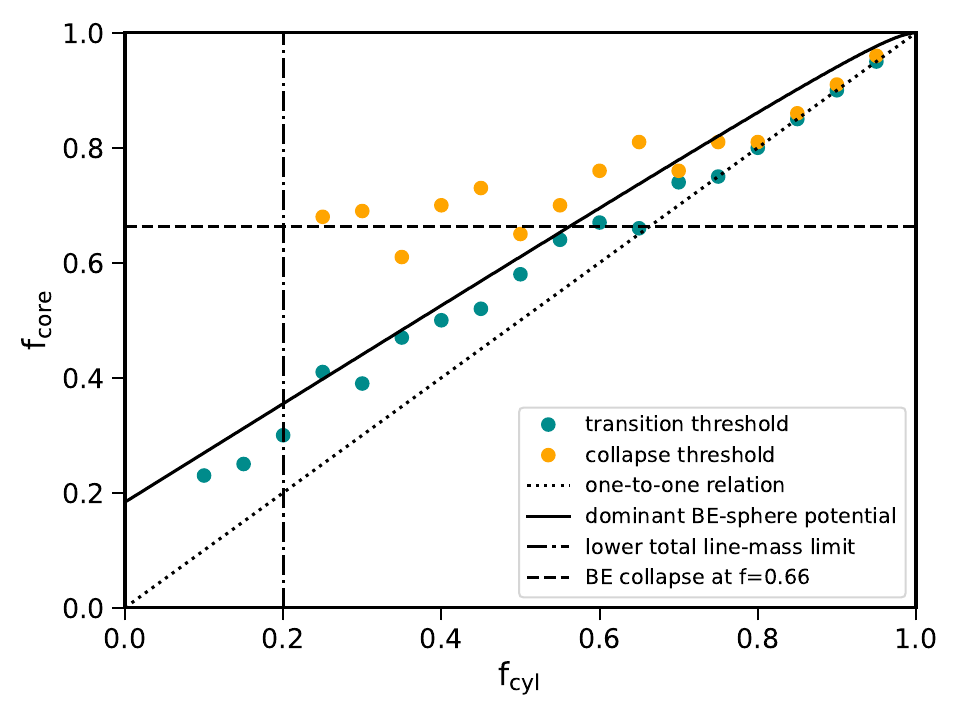}
    \caption{Threshold line-masses at the position of the core in dependence of the
    filament line-mass $\fcyl$. The begin of the non-linear evolution in acceleration
    is given by the cyan points for the velocity by the orange points. The solid line
    indicates where the Bonnor-Ebert sphere potential dominates over the filament
    potential. The vertical dashes-dotted line shows the lower mass limit for a
    collapsing Bonnor-Ebert sphere and the horizontal dashed line shows the result
    of calculating the maximum supported Bonnor-Ebert mass assuming a external
    pressure given by the density within the filament.}
    \label{fig:fnonlin}
  \end{figure}

  We plot the measured threshold values for all filaments in Fig.~\ref{fig:fnonlin},
  where the x-axis shows the overall line-mass of the filament and the y-axis shows
  the line-mass at the core position at the transition to the respective phase, cyan
  for the threshold in acceleration and orange for the threshold in velocity. While
  the threshold in acceleration is line-mass dependent, the threshold in velocity is
  very similar for lower filament line-masses. Only for larger filament line-masses,
  both thresholds set in very early and therefore follow the one-to-one relation shown
  as the dotted line. Cores forming in filaments with very low line-masses of
  $\fcyl=0.2$ and below do not collapse in our simulations, consistent with the lower
  Bonnor-Ebert mass limit as calculated by \citet{fischera2012} which we indicate by
  the dashed-dotted vertical line. While the thresholds seem to be fitted well from
  visual inspection, we only performed a single simulation per line-mass. A more
  rigorous statistical approach would be to perform several simulations for the same
  line-mass and determine mean values with respective error bars. However, the
  fitted results seem to be relatively robust with respect to the number of outputs
  we use to smooth the data and the error threshold we use to determine the begin
  of the non-linear deviation.

  The first threshold seen in acceleration and density can be explained by the
  change of the main gravitational potential. As the core grows, its gravitational
  potential changes from being dominated by the overall filament potential to being
  dominated by the Bonnor-Ebert sphere potential. This leads to a second order effect
  where the core is stronger concentrated to its centre. Therefore, the perturbation
  grows faster than expected. We can determine the expected transition line-mass at
  the core position by assuming that the Bonnor-Ebert sphere potential of the
  perturbation should be deeper than the overall filament potential even for the
  cores largest extent which is given by the unperturbed filament radius $R$:
  \begin{equation}
    \Phi_\text{BE}(R) > \Phi_\text{cyl}(R)
    \label{eq:phi}
  \end{equation}
  We calculate the potential of the filament as well as the Bonnor-Ebert sphere by
  computing the numerical solution to Eq.~\ref{eq:fil} and Eq.~\ref{eq:le} using the
  explicit 5th-order Runge-Kutta method built into the SciPy integrate package. For
  the same central density, the filament potential is deeper than the Bonnor-Ebert
  sphere potential at all radii. We then iteratively determine how large the central
  density of the Bonnor-Ebert sphere has to be in order to fulfill the condition given
  by Eq.~\ref{eq:phi} and use it to calculate the corresponding line-mass value
  $\text{f}_\text{core}$ given by Eq.~\ref{eq:rho}. The result is shown in
  Fig.~\ref{fig:fnonlin} as the solid line. As one can see, the cyan values follow
  this prediction closely except for higher line-masses than $\fcyl=0.65$ where
  the deviation seems to be instanteneous.

  Likewise, the threshold in velocity can be explained by assuming that it is the
  Bonnor-Ebert sphere that also determines the collapse of the core. As the maximum
  supported mass of a Bonnor-Ebert sphere in Eq.~\ref{eq:Mbe} is given by the
  external pressure, one can estimate at which line-mass threshold the mass within
  a core exceeds this maximum value. However, the external pressure is not simply
  given by the external medium around the filament but is considerably larger as
  the core sits within the filament itself. This means we have to determine the total
  mass of the core as well as the average pressure on its surface. However, this
  is not trivial as a core located within a filament is typically neither spherical
  nor does it have a clearly defined surface. We therefore make use of some
  simplifications. On the one hand, we estimate the mass of a core by numerically
  integrating the mass of a spherical region centred inside an unperturbed
  filamentary density distribution as given by Eq.~\ref{eq:rho}:
  \begin{equation}
    M_\text{core} = \int_0^R\int_0^\pi\int_0^{2\pi}
    \frac{\rho_c r^2\sin(\theta)\,dr\,d\theta\,d\phi}
    {\left(1+r^2\left(1-\cos(\theta)^2\right)/H^2\right)^2}
  \end{equation}
  where we expressed the cylindrical radius in spherical coordinates $r$, $\theta$
  and $\phi$. On the other hand, we estimate the average external pressure at the
  surface of this region via the density which we get by solving the surface integral
  numerically and normalising by the surface area:
  \begin{equation}
    \langle\rho_\text{BE}\rangle = \frac{1}{4\pi R^2}\int_0^\pi\int_0^{2\pi}\frac{\rho_c R^2\sin(\theta)\,d\theta\,d\phi}
    {\left(1+R^2\left(1-\cos(\theta)^2\right)/H^2\right)^2}
  \end{equation}
  For the radius we use the maximum volume a core can occupy. Generally, this value
  is given by the filament radius. However, for large line-masses the dominant
  wavelength becomes smaller than the filament diameter. This means the core radius can
  at maximum be half the distance to the next core. We therefore define the core radius
  as the minimum value of this value and the unperturbed filament radius. These
  simplifications neglect that a core represents an overdensity within the filament.
  Thus, the density at the surface and mass within the core should be larger than
  for the initial filament. However, this effect is somewhat counterbalanced as
  both the mass within the core as well as the density at its surface should increase
  in an analogous manner for self-similar profiles like the Bonnor-Ebert sphere.

  \begin{figure}
    \centering
    \includegraphics[width=1.0\columnwidth]{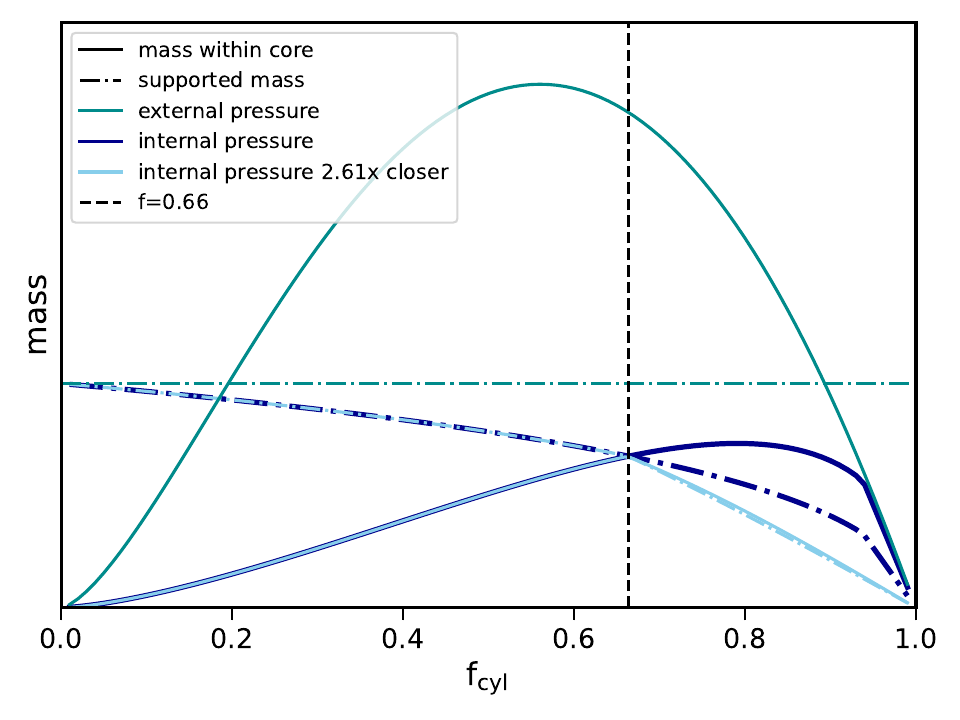}
    \caption{Comparison of the mass within a core, given by the solid lines, to the
    supported Bonnor-Ebert mass, given by the dashed-dotted lines, as function of the
    line-mass. The cyan lines show the model of only taking the external pressure into
    account where the core mass is given by the total mass of the dominant wavelength.
    The dark blue lines show our model of a spherical core located within a filamentary
    density profile where cores become unstable for line-masses above $\fcyl=0.66$ as
    marked by the vertical dashed line. The light blue lines show the closest separation
    the cores can have while still being unstable which is 2.61 times closer than the
    dominant wavelength.}
    \label{fig:integration}
  \end{figure}

  We plot both masses as function of the line-mass in Fig.~\ref{fig:integration}, where
  the integrated mass is given by the dark blue solid line and the maximum supported
  mass calculated by using the average surface density in Eq.~\ref{eq:Mbe} is given
  as dark blue dashed-dotted line. We did not include a y-axis as all masses scale
  the same with respect to the external pressure. Therefore, the calculation is scale-free.
  As one can see, the mass within the core surpasses the supported Bonnor-Ebert mass
  at $\fcyl=0.66$, indicated as dashed vertical line, which is exactly the value we
  plot as dashed horizontal line in Fig.~\ref{fig:fnonlin}. Thus, as soon as the
  perturbation reaches this threshold, collapse will set in. If the initial line-mass
  is already above this value, any perturbation will lead to a collapsing core. This is
  the reason why we observe a direct transition to the exponential increase in
  Fig.~\ref{fig:fnonlin} for line-mass values above $\fcyl=0.60$. For very large
  line-masses, one can see a break in the curves as the core distances on the dominant
  perturbation wavelength become smaller than the filament diameter.

  As a comparison, we also plot the total mass contained in the dominant wavelength
  compared to the Bonnor-Ebert mass supported by the external pressure as gray solid
  and dashed-dotted lines. The intersections show the calculation of \citet{fischera2012}
  where only line-masses of values $0.20 < \fcyl < 0.89$ form collapsing cores. Despite
  approaching the same value for low line-masses, the internal pressure reduces the
  supported mass to smaller and smaller values compared to the external pressure as
  the line-mass grows. This is due to the density profile, which is flat at low line-masses
  and increasingly steeper at larger line-masses. Here, one can also see that the constraint
  on the core radius as described above, limits the mass within the core to the maximum
  mass within the dominant wavelength. Therefore, the solid dark blue line follows
  the solid gray line. However, this does not influence the collapse as both masses, the
  core mass as well as the supported mass, fall off steeply. Thus, the internal pressure
  can also demonstrate why cores forming within high line-mass filaments are even able to
  collapse which cannot be explained by the external pressure as shown by the cyan lines.
  It is not the total mass of the dominant wavelength determining the collapse, but the
  local Bonnor-Ebert mass within the filament.

  \subsection{Effects on collapsing core separation}

  \begin{figure}
    \centering
    \includegraphics[width=1.0\columnwidth]{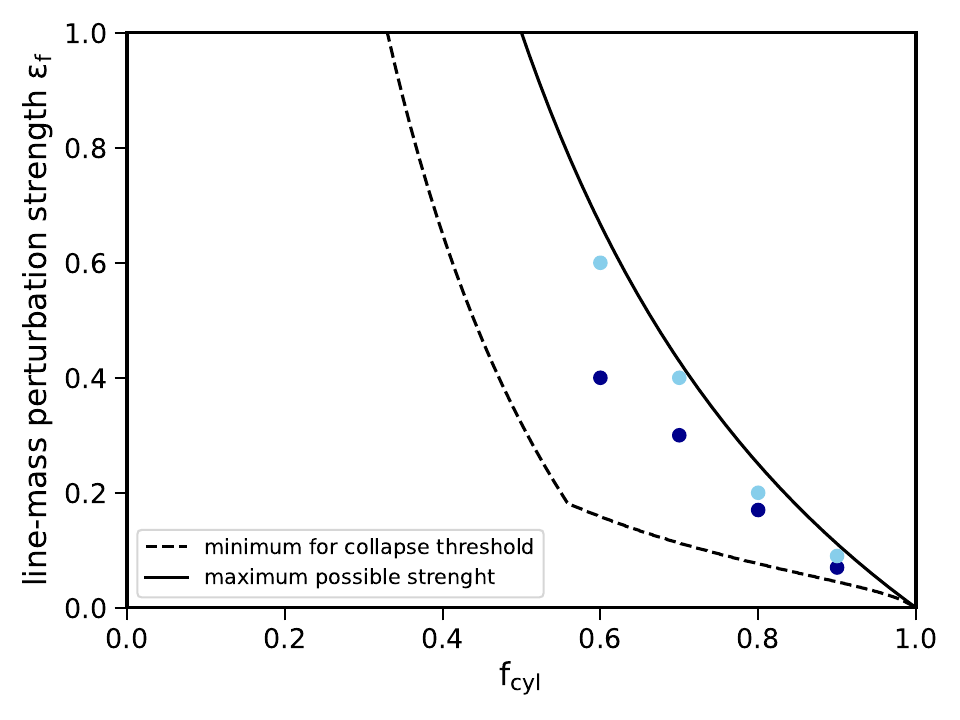}
    \caption{Minimum and maximum line-mass perturbation strength for direct collapse
    as function of the filament line-mass. The minimum is given by the fact that the
    Bonnor-Ebert sphere potential has to dominate over the filament potential and at
    the same time the line-mass maximum has to be above $\fcyl=0.66$. The maximum is
    given by the fact that the line-mass cannot exceed a value of 1.0 in equilibrium.
    The light and dark blue dots show the perturbation strength we used for our test
    simulations given in Fig.~\ref{fig:fwhm}.}
    \label{fig:strength}
  \end{figure}

  If the collapse is thus given by a local Bonnor-Ebert criterion, the question
  arises if stars can form on any given core separation. However, there is an
  upper limit on how close unstable cores can form given by the maximum size they occupy.
  Due to our constraint that the core radius is given by the minimum of the filament
  radius and half the distance to the next core, the mass within the core reduces
  faster than the average ambient density for shorter and shorter core separations.
  The transition point where one cannot find any unstable cores anymore is shown
  by the light blue lines in Fig.~\ref{fig:integration} which is the case for cores
  which are spaced 2.61 times closer than the dominant wavelength. For this value,
  cores are also spaced closer than the critical wavelength. This means while small
  perturbation will not be able to grow into unstable cores, large line-mass perturbations
  which exceed the collapse threshold of $\fcyl=0.66$ and at the same time lead to a
  dominant Bonnor-Ebert sphere potential should be able to collapse. We plot this
  limit in Fig.~\ref{fig:strength} in terms of minimum line-mass perturbation
  strength as dashed black line. We also include the upper limit of a line-mass of
  $\fcyl=1.0$ as solid black line.

  We test this prediction by simulating high line-mass filaments with varying
  perturbation wavelengths and compare if the cores collapse on the wavelength
  of the dominant mode or on the separation given by the initial condition. In
  contrast to the former simulations which used a perturbation in density, here
  we use a perturbation in line-mass. This is due to the fact that a large change
  in line-mass cannot be adequately reproduced with a linear perturbation in
  density as given by Eq.~\ref{eq:eps}. For large changes in line-mass as shown
  in Fig.~\ref{fig:strength}, the linear perturbation strength $\epsilon$ even
  exceeds values of 1.0 and the resulting initial condition would not be a
  continuous filament. Therefore, a large line-mass perturbation means that our
  simulation already begins in the non-linear regime where we cannot derive a
  self-consistent initial condition in velocity. Nevertheless, we are only
  interested if large line-mass perturbations can lead to core collapse on
  shorter wavelengths than allowed by linear growth. This is also exactly the
  initial condition which arises when filaments form in molecular clouds with
  large variations in line-mass. Thus, we neglect any initial velocity along the
  filament and set-up the density profile consistent with Eq.~\ref{eq:contrast}.

  \begin{figure}
    \centering
    \includegraphics[width=1.0\columnwidth]{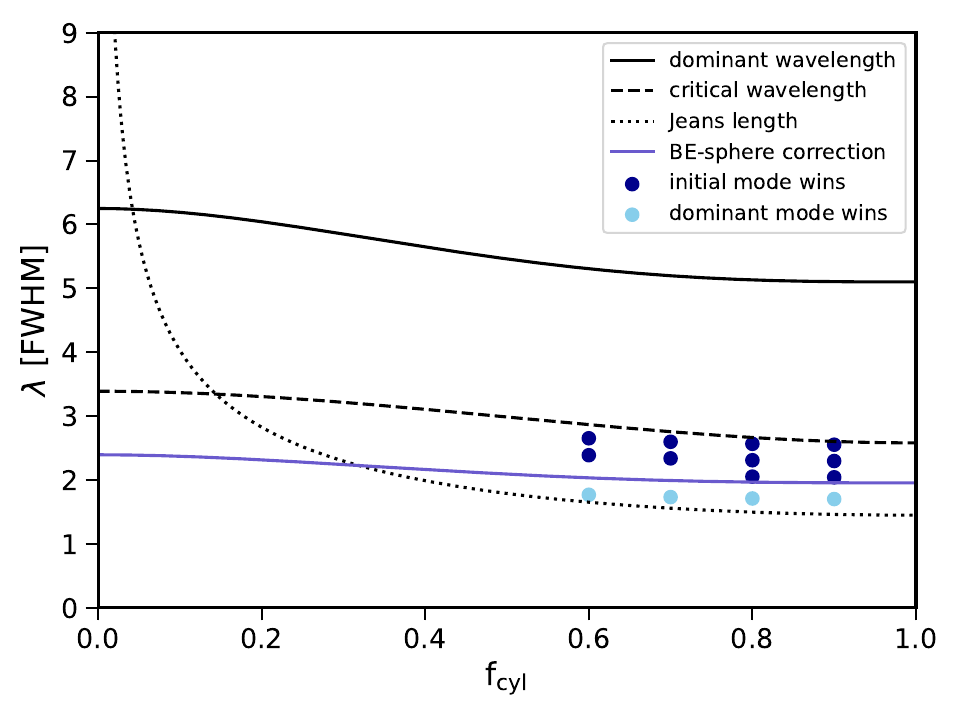}
    \caption{The Bonnor-Ebert sphere collapse threshold, shown as blue solid line,
    in comparison to the dominant and critical wavelength, given by the black solid and
    dashed lines, respectively. The wavelengths are given in units of the full width at
    half maximum of the filament profile. We assess the threshold by performing test
    simulations indicated by markers where the colour indicates if the initial perturbation
    is preserved during collapse, given by the dark blue markers, or if it is lost and
    a collapsing core forms on the dominant wavelength, as shown by the light blue markers.}
    \label{fig:fwhm}
  \end{figure}

  We show the test cases and their outcome in Fig.~\ref{fig:fwhm} where the
  dominant and critical wavelength are given in units of the full width at half
  maximum (FWHM) of the filament profile as defined in \citep{fischera2012}. The
  dominant wavelength is shown by the solid black line and the critical wavelength,
  about a factor of two shorter, as dashed black line, respectively. We also plot
  the three-dimensional Jeans length defined by the central density
  \begin{equation}
    \lambda_J = \sqrt{\frac{\pi c_s^2}{G\rho_c}}
  \end{equation}
  as dotted black line. The Bonnor-Ebert sphere correction of a factor 2.61
  shorter than the dominant wavelength is indicated by the blue solid line.
  We simulate filaments with perturbation lengths towards longer and shorter
  separations compared to this value by setting up line-mass perturbations
  with one third and half the dominant wavelength. In addition, we also set up
  perturbations which are 2.22 and 2.5 times shorter than the dominant wavelength
  by doubling the box size reducing the box size by a factor of 0.9 and 0.8.
  If only one collapsing core forms, the wavelength on the scale of the box
  size was able to grow faster and therefore the dominant wavelength is the
  fastest growing mode. The dark blue dots indicate simulations where all cores
  collapse on the initial perturbation length and the light blue dots show
  simulations where only one core collapses.

  \begin{figure}
    \centering
    \includegraphics[width=1.0\columnwidth]{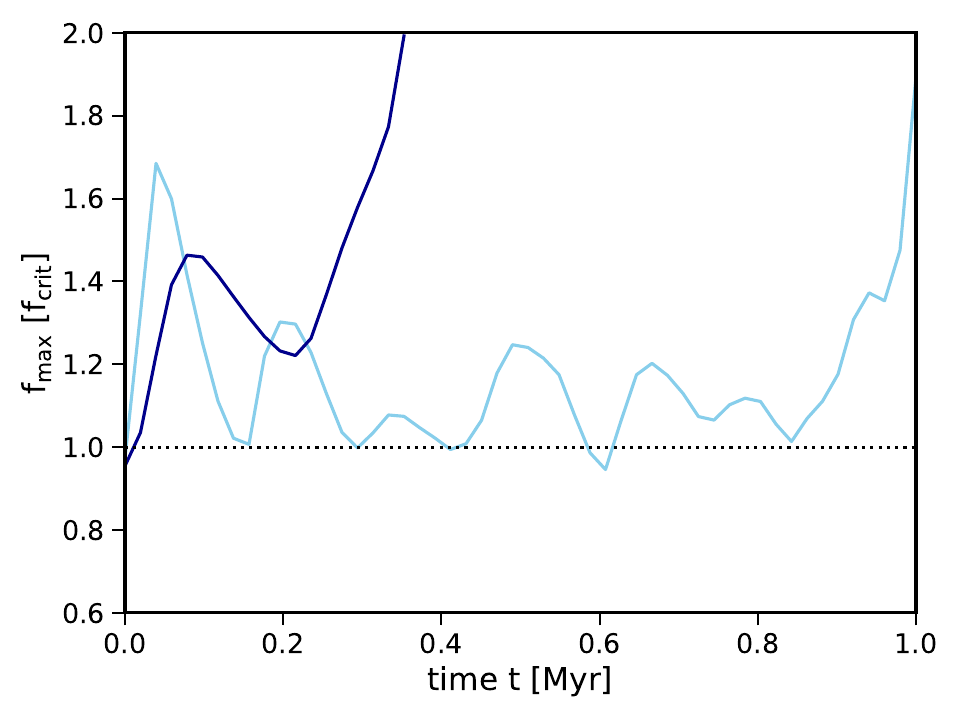}
    \caption{Evolution of the maximum line-mass in the simulation of a
    filament of $\fcyl=0.9$ for a perturbation wavelength of three times
    shorter, shown in light blue, and 2.22 times shorter, shown in dark
    blue, than the dominant wavelength. In the former case the core does
    not collapse directly but the line-mass oscillates with values above
    the critical line-mass until eventually a single collapsing core
    forms on the dominant wavelength.}
    \label{fig:osci}
  \end{figure}

  As one can see, all of our test cases with wavelengths above but close to
  the Bonnor-Ebert collapse threshold are able to trigger direct collapse of
  cores even if their separation lies below the critical wavelength. However,
  modes of one third of the dominant wavelength do not collapse directly in
  our simulations but form a collapsing core on the dominant wavelength. Both
  evolutions, the collapse of the initial mode and of the dominant mode, are
  shown in Fig.~\ref{fig:osci} for a filament with $\fcyl=0.9$. Here we plot
  the evolution of the maximum line-mass within the filament as function of
  time. The light blue line shows the evolution of the mode three times shorter
  and the dark blue line shows the evolution of the mode 2.22 times shorter
  than the dominant wavelength. While the cores for the dark blue case rapidly
  grow and collapse in a very short time, the cores for the light blue case
  show a soundwave oscillation despite having line-masses which consistently
  exceed the critical value of where the filament should be radially stable.
  There is no collapse until a larger core grows on the dominant wavelength
  which eventually collapses after one Myr. This demonstrates that core
  collapse is indeed not given by the loss of radial stability but by the
  mass exceeding the local collapse threshold.

  In order to make sure that it is not the perturbation strength which is
  preventing the collapse in the cases of where the dominant mode wins,
  we set it close to the maximum possible value where the perturbed initial
  filament line-mass approaches 1.0. This is visualised in
  Fig.~\ref{fig:strength} as light blue dots. In contrast, the initial
  perturbation strength for the simulations where the initial mode collapses
  directly is shown as dark blue dots. While we expect that cores should be
  able to directly collapse if the perturbation strength is above the dashed
  threshold value, we notice that low initial strengths close to this threshold
  are not able to overcome the restoring forces leading to oscillating
  soundwaves. The reason for this is likely due to the initial conditions.
  On the one hand, the threshold value of $\fcyl=0.66$ was measured by the
  growth of linear perturbations. As we only use an initial perturbation
  in line-mass, the missing inflow velocity along the filament could be
  increasing the perturbation strength needed for direct collapse. On the
  other hand, the initial perturbation is not set-up as a Bonnor-Ebert
  sphere. This means the overdensity needs time to adjust to a Bonnor-Ebert
  sphere profile in order to collapse. As this happens on the timescale of
  a sound crossing time, the filament first enters a phase of oscillating
  overdensities. We do observe this effect in some simulations with low
  initial perturbation strength where there is a long period of oscillating
  soundwaves before the cores eventually do collapse on the initial mode.

%%%%%%%%%%%%%%%%%%%%%%%%%%%%%%%%%%%%%%%%%%%%%%%%%%%%%%%%%%%%%%
\section{Conclusions and discussion}
  \label{sec:conclusion}

  We have shown that if cores form with enough mass, they are able to
  collapse even if they lie closer together than the minimum wavelength
  for core growth given by linear perturbation theory. Our results help
  to explain why some observations show filaments with star-forming cores
  which are separated closer than the dominant wavelength. Instead of collapsing
  due to linear growth, these cores could have formed as large line-mass
  perturbations during filament formation. Moreover, they could also explain
  why some observations show a hierarchical fragmentation. Low line-mass filaments
  should form dense, high line-mass perturbations on the dominant scale. As
  soon as it reaches the line-mass threshold for direct collapse, any left-over
  overdensity within these regions is highly susceptible to collapse on very
  short separations. While our results do not allow the collapse on the exact value
  of the Jeans length set by the central density, our derived wavelength threshold
  is very similar and would be hard to distinguish observationally. In summary,
  our main conclusions are the following:
  \begin{itemize}
    \item Perturbation growth in filaments shows non-linear effects as soon as the
      Bonnor-Ebert sphere gravitationally dominates over the filament potential. As
      a result, the profile becomes more concentrated and therefore the density
      evolves faster than predicted by linear growth.
    \item Core collapse, as indicated by the non-linear evolution of the velocity
      along the filament axis, sets in before the perturbation reaches the
      critical line-mass. This means core collapse is not triggered by the loss of
      radial stability of the filament.
    \item We can explain the early collapse by assuming the mass within the core
      has reached the Bonnor-Ebert mass limit given by the pressure within the
      filament. This sets the collapse threshold to a line-mass value of $\fcyl=0.66$.
    \item For large enough line-mass perturbations, cores therefore can go into
      direct collapse even if their separation is closer than the minimum wavelength
      required for perturbation growth. We predict that this separation can nominally
      be 2.61 times closer than the dominant wavelength.
  \end{itemize}
  The main caveat of our study is that we do not measure the collapse threshold of
  $\fcyl=0.66$ with high accuracy. From a theoretical standpoint, one could also argue
  for a collapse threshold of $\fcyl=0.73$ as this is the line-mass when the density
  contrast according to Eq.~\ref{eq:contrast} from central to boundary density within
  a filament reaches the value required for collapse of a Bonnor-Ebert sphere of 14.1.
  We test this possibility by measuring if the contrast of central to boundary density
  actually reach this value in our simulations before the cores collapse. However, this
  does not seem to be the case. For initial filament line-masses below $\fcyl=0.73$ the
  core collapse begins before the density contrast reaches a value of 14.1. Therefore,
  it is indeed a mass threshold and not a density contrast which triggers the begin of
  collapse. Nevertheless, a slight shift to larger line-mass values in the collapse
  threshold also only leads to a minor shift to larger values of the minimum wavelength
  for direct collapse.

  A central question for our scenario is if filaments form with large enough line-mass
  perturbations to trigger direct collapse. Observationally, it is not easy to distinguish
  overdensities which have formed with high line-mass from those which have grown to
  large values. However, detected line-mass deviations in filament are quite large.
  For example, \citet{roy2015} find a standard deviation in line-masses of ~7\% which
  locally also reaches values of up to 30\%. Considering that most observed filaments
  are also rather large in line-mass, the perturbation strengths required to go into
  direct collapse are not that high as can be seen in Fig.~\ref{fig:strength}.

%%%%%%%%%%%%%%%%%%%%%%%%%%%%%%%%%%%%%%%%%%%%%%%%%%%%%%%%%%%%%%
\begin{acknowledgements}
   This research was supported by the Excellence Cluster ORIGINS which is funded by the
   Deutsche Forschungsgemeinschaft (DFG, German Research Foundation) under Germany’s
   Excellence Strategy - EXC-2094 - 390783311.
\end{acknowledgements}

%%%%%%%%%%%%%%%%%%%%%%%%%%%%%%%%%%%%%%%%%%%%%%%%%%%%%%%%%%%%%%
\bibliographystyle{aa}
\bibliography{collapse}

%%%%%%%%%%%%%%%%%%%%%%%%%%%%%%%%%%%%%%%%%%%%%%%%%%%%%%%%%%%%%%%
%\begin{appendix}
%\end{appendix}

\end{document}